\documentstyle[12pt] {article}
\begin{document}
\thispagestyle{empty}

\title{Charge Symmetry Breaking in the Valence Quark Distributions
of the Nucleon}
\author{C.J. Benesh$^{1,2,3}$ and T. Goldman$^{1}$,\\
1-Los Alamos National Laboratory, Los Alamos, NM  87545,\\
2-International Institute for Theoretical and Applied Physics,\\ Iowa
State University, Ames, Iowa 50010,\\
3-Current Address: Nuclear Theory Center, Indiana University,\\
 Bloomington, IN 47408.}

\maketitle
\setcounter{page}{0}

\begin{abstract}
Using a quark model, we study the effect of charge symmetry
breaking on the valence quark distributions of the nucleon. The effect
due to quark mass differences and the Coulomb interaction of the
electrically charged quarks is calculated and, in contrast to recent
claims, found to be small. In addition, we investigate the effect of
charge symmetry breaking in the confining interaction, and in the
perturbative evolution equations used to relate the quark model
distributions to experiment. We find that both these effects are small,
and that the strong charge symmetry breaking effect included in the
scalar confining interactions may be distinguishable from that
generated by quark mass differences. 	
\end{abstract}

\pagebreak

\section{Introduction}
	
The study of charge symmetry and its breaking is almost as old as
nuclear physics itself. From the earliest days of isospin\cite{1}, to
modern attempts to understand small effects in the nucleon-nucleon
interaction\cite{2}, the study of this symmetry has provided a rare
window into the non-perturbative dynamics of low energy hadronic
phenomena. The interactions responsible for charge symmetry breaking
(CSB) are largely understood and relatively weak, so that the study of
CSB provides a sensitive filter with which to test the hadronic
wavefunctions of nuclei and nucleons.  In the case of QCD, where the
interactions that bind quarks into hadrons are only understood
schematically and the theoretical landscape is cluttered with different
phenomenological models, such probes may prove especially valuable.

In this paper, we study the effect of the breaking of charge symmetry
on the valence quark distributions of the nucleon. Our interest in this
topic is generated by the observation of unexpected effects in both the
sea\cite{3} and spin-dependent\cite{4} quark distributions of the
nucleon, and by the possibility that large CSB effects in the valence
distributions of the nucleon may play a non-negligible role in the
extraction of $\sin^2\theta_W$ from $\nu-N$ data\cite{5}. Recent
calculations have claimed that charge symmetry breaking may be greatly
enhanced by kinematic effects associated with diquarks in the nucleon
wavefunction\cite{6}, and that the resulting CSB may be large enough to
be observed directly. Our calculations provide an important check on
the model dependence of these results, as well as investigating several
sources of CSB not considered in Ref. 6.

The paper is structured as follows: In Section 2, the method we use to
extract quark distributions from quark models is reviewed, with
particular emphasis on those details of the calculation that will be
affected by charge symmetry breaking. In the next section, the shift in
the valence quark distributions due to the breaking of charge symmetry
by quark masses, and by Coulomb effects is calculated for both the
minority and majority quark distributions in the nucleon. The effect of
the difference between the neutron and proton mass is also discussed
and, after concluding that it contains no physics, discarded. The
fourth section is devoted to a discussion of the possibilty of charge
symmetry breaking in the confining interaction itself, and the
possibility that such effects may be distinguishable from ordinary
quark mass effects on the valence distributions, while in the fifth
we calculate the charge symmetry breaking effect due to the perturbative
evolution of the valence distributions that is required if the quark
model is to make contact with high energy data. The final section
compares our results to those of references 5 and 6, and discusses the
prospects for measuring CSB in the valence distributions directly.

\section{ Valence Quark Distributions in the Los Alamos Model Potential
(LAMP).}

Quark models were originally constructed to provide a description of
low energy hadronic data using only effective interactions between
valence quarks. In order to make sensible calculations of valence quark
distributions, the stark simplicity of quark models must be reconciled
with the richer vision afforded by data from higher energies, where
nucleons are composed not only of valence quarks, but also of sea
quarks and gluons. Remarkably, these two very different pictures may
be accomodated by QCD via the renormalization group\cite{7}. Quark
models may be interpreted as representations of QCD at an intermediate
renormalization scale, $\mu_{QM}^2$, large enough so that the quark
substructure of the nucleon is revealed, but small enough that the sea
quarks and glue are almost entirely absorbed by a redefinition of the
valence quarks. Parton distributions at this intermediate scale may be
calculated in terms of the quark model, and then evolved to higher
energies using perturbative QCD\cite{8} and compared to data.

The first step in this procedure is to calculate the quark
distributions at the quark model scale. For unpolarized scattering, the
relevant matrix elements are given by\cite{8}
\begin{eqnarray}
q_i(x) &=& \, {1\over 4\pi} \int  d\xi^-  \, e^{iq^{+}\xi^-}
<N\vert\bar\psi_i(\xi^-)\gamma^+\psi_i(0)
\vert N>\vert_{LC} \nonumber\\
\bar q_i(x) &=&-{1\over 4\pi} \int  d\xi^- \, e^{iq^+\xi^-}
<N\vert\bar\psi_i(0)\gamma^+\psi_i(\xi^-)\vert N>\vert_{LC},
\end{eqnarray}
where $q^+=-Mx/\sqrt{2}$ (with $x\equiv x_{Bj}$ the Bjorken scaling
variable), $\psi_i(\bar\psi_i)$ are field operators for quarks of
flavor $i$, $\gamma^+$ is a Dirac gamma matrix, and the subscript LC
indicates a light cone condition on $\xi$, namely that
$\xi^+=\vec\xi_\perp=0$.

For a given model, valence quark distributions can be obtained via a
number of different prescriptions that have arisen in the
literature\cite{9,10,11}. The approach we adopt consists of a
straightforward evaluation of the matrix elements of Eq. 1 in a
Peierls-Yoccoz projected momentum eigenstate, assuming that the time
dependence of the field operator is dominated by the lowest eigenvalue
of the Dirac equation used to obtain the wavefunctions of the struck
quark. The details of this procedure are described in Ref. 9, where the
valence quark distributions are shown to be given by
\begin{equation}
xq_V^i(x)= {MxN_i\over \pi V}\left \{\left [ \int_{\vert k_-\vert}^\infty dk
\, 
G_i(k)
\left (t_{i0}^2(k) + t_{i1}^2(k) + 2{k_-\over k}t_{i0}(k)t_{i1}(k)\right
)\right ] +\left [ k_- \rightarrow k_+\right ]\right \},
\end{equation}
where 
\begin{eqnarray}
G_i(k) &=& \int r dr \sin kr \Delta_{s1}(r) \Delta_{s2}(r) EB(r), \nonumber
\\
V &=& \int r^2 dr \Delta_i(r) \Delta_{s1}(r) \Delta_{s2}(r)
EB(r),\nonumber\\
t_{\alpha 0}(k)& =& \int r^2 dr j_0(kr) u_\alpha(r),\nonumber\\
t_{\alpha 1}(k)& =& \int r^2 dr j_1(kr) v_\alpha(r),\nonumber\\
\Delta_\alpha(r)&=& \int d^3z \phi^\dagger_{0\alpha}({\bf z}-{\bf r})
\phi_{0\alpha}({\bf r}),
\end{eqnarray}
with $\phi_{0\alpha}({\bf r})$ the ground state valence quark
wavefunction for a quark of flavor $\alpha$, with upper and lower
components $u_\alpha(r)$ and $i{\bf\sigma\cdot r}v_\alpha(r)/r$,
$k_\pm= \omega_i \pm Mx$, with $\omega_i$ the ground state struck quark
energy eigenvalue, $EB(r) =~~<EB, {\bf R_{CM}}={\bf r}|EB, {\bf
R_{CM}}={\bf 0}>$ is the overlap function for two ``empty bags''
separated by a distance $r$, which accounts for the dynamics of the
confining degrees of freedom.  Finally, the subscripts $s1$ and $s2$
denote the flavor of the two spectator valence quarks that make up the
nucleon.

All calculations described in this paper are carried out using the Los
Alamos Model Potential (LAMP)\cite{12}, in which valence quarks are
confined by a linear potential of the form
\begin{equation}
V(r)=\beta k(r-r_0)
\end{equation}
with parameters $k$=0.9 GeV and $r_0$=0.57 fm chosen to reproduce the
average nucleon-delta mass, and $\beta$ is a Dirac gamma matrix.  We
further assume that the function $EB(r)$ is a constant\cite{12a}, and
unless otherwise noted, quarks are taken to be massless.

\section{Charge Symmetry Breaking} 
  
As we have already noted, the sources of charge symmetry breaking are
light quark mass differences and the electro-magnetic interaction. CSB
effects may be manifested either directly, as a result of explicit mass
or interaction terms in the quark model hamiltonian, or indirectly as a
result of mixing between the symmetry violating terms and the strong
interaction. In this section, we calculate the direct terms, deferring
discussion of the numerically smaller mixed interactions for later.

At the partonic level, charge symmetry predicts that the u quark
distribution in the proton is equal to the d quark distribution in the
neutron, with a corresponding prediction for the d distribution.  A
measure of the extent to which the symmetry is broken is given by the
ratios
\begin{eqnarray}
R_{CSB}^{maj}(x) &=& 2{\left [ u^p_V(x)-d^n_V(x)\right ]}\over{\left [ u^p_V(x)+d^n_V(x)\right ]}\nonumber\\
R_{CSB}^{min}(x) &=& 2{\left [ d^p_V(x)-u^n_V(x)\right ]}\over {\left [ d^p_V(x)+u^n_V(x)\right ]},
\end{eqnarray}
where $u(d)^{p(n)}_V(x)$ denotes the up(down) valence quark
distribution in the proton(neutron). 
 
\subsection{CSB in Quark Wavefunctions}

The simplest mechanism for altering the shape of the valence
distributions of the nucleon in Eq. 2 is to change the model
wavefunctions for the struck and spectator quarks. Since these
wavefunctions are generated by solving a Dirac equation in the mean
confining field of the other quarks, direct CSB effects may be included
without additional assumption by including appropriate terms in the
Dirac Hamiltonian.
	
We have recalculated the valence quark wavefunctions assuming quark
masses $m_u$= 4 MeV and $m_d$= 8 MeV\cite{13}. In Figs. 1 and 2, the
CSB ratios produced by these wavefunctions are  indicated by the dashed
curves. For minority quarks, the ratio starts small and rises to +1.5\%
at large $x$, while the majority quark ratio rises more slowly, but is
still positive.  The fact that these ratios are of the same sign and
comparable in magnitude may be understood, since, when the struck quark
is a majority quark, the two spectators are in a charge symmetric
state. The only change is due to the change in the valence quark
wavefunction. When a minority quark is struck, the change in the
wavefunction of the struck quark is compensated for by the change in
the wavefunctions of the two spectators, which alter the momentum
projection factor $G_i(k)$ in Eq. 1, and are of opposite sign and
roughly twice as large as the effect produced by the struck quark
wavefunction.

In addition, we have calculated the effect of the Coulomb interaction
between the charged quarks using a mean field approximation for the
electric potential between a quark of charge $q$ and the other valence
quarks, given by
\begin{equation}
V_{Coul}(r)= \alpha q(Q_N -q)\int d^3r^\prime
 {\phi^\dagger(r^\prime)\phi(r^\prime)\over \vert {\bf r}- {\bf r^\prime}\vert},
\end{equation}
where $Q_N$ is the charge of the nucleon being studied. The CSB
contribution generated by the Coulomb effect on the quark wavefunctions
is shown by the dot-dashed curves in Figs. 1 and 2. Again, the CSB
effect produced by perturbing the wavefunctions is significantly
smaller for the minority quark distributions. In this instance, the
change in the wavefunction due to the Coulomb force is, to first order
in $\alpha$, the same for either minority quark. Even for the majority
quarks, the Coulomb correction is much smaller than 1\%. Magnetic
corrections have not been calculated explicitly, but since they are
similar in structure to the color magnetic corrections responsible for
breaking SU(4) symmetry in the quark distributions\cite{9}, yet are
suppressed relative to those corrections by both the smallness of the
electromagnetic coupling constant and by color SU(3) factors, their
effect should also be very small.

\subsection{CSB in the Dirac Eigenvalue}

Along with the wavefunctions, solution of Dirac equation provides an
energy eigenvalue which determines the time dependence of the lowest
mode of the confined quark field. This energy is just that required to
break the bonds which tie the struck quark to the spectators, and as
such is sensitive to the nature of the interactions that bind the
quarks together. Since the dependence of the quark distribution
functions on this eigenvalue has the same functional form regardless of
whether the changes in the eigenvalue are produced by quark mass
effects or Coulomb interactions, we have simply taken the changes in
the eigenvalues produced by the Hamiltonian changes already described
and added them together. For the proton the quark eigenvalues were
shifted from the value for massless quarks (361.8 MeV) to
$\omega_u=364.2$ MeV and $\omega_d=365.0$ MeV.  The corresponding
values for the neutron are $\omega_u=363.0$ MeV and $\omega_d=365.6$
MeV.  The CSB ratios obtained using these eigenvalues, and massless
quark wavefunctions, are shown by the solid lines in Figs. 1 and 2. At
large $x$, these shifts produce CSB effects on the order of 2-3\%,
which, may be understood from Eq. 1 as the fractional shift in the
quark eigenvalue enhanced by the slope of the unperturbed quark
distribution. (This is the closest analog to the diquark mass shift of
Ref. 6, and correspondingly produces the largest effect.)

\subsection{Proton-Neutron Mass Difference}

The results of changing the nucleon mass parameter appearing in Eq. 1
to reflect the difference in mass of the proton and neutron is shown by
the dotted curves in Figs. 1 and 2. Like the quark eigenvalue, the
effect of changing the nucleon mass parameter is enhanced, at large
$x_{Bj}$, by the slope of the unperturbed quark distribution function.

This effect, however, will not be present if the data is analyzed in
the conventional manner because the nucleon mass appears not only as an
explicit parameter in Eq. 1, but is also implicit in the definition of
$x_{Bj}$ (In deep inelastic leptoproduction, for example,
$x_{Bj}=Q^2/2M_{nuc} q_0$ in the target rest frame.). In fact, the
combination $M_{nuc}x_{Bj}$, which is all that appears in Eq. 2 for
$xq(x)$, is completely independent of $M_{nuc}$!  This is in accord
with what one would expect in a light cone formalism, where $P^+$ is a
kinematical variable, and therefore immune to the dynamical effects
which violate charge symmetry.  (While it is certainly possible to
alter the usual analysis by rescaling $x_{Bj}$ once the neutron
structure function is extracted from the raw data, it not clear what
would be learned by comparing the probability of finding quarks in the
proton at one momentum to that in the neutron at a slightly different
momentum.) Hence, we assert that there is no CSB effect in the parton
distributions due to the neutron-proton mass difference\cite{14}.
 
\section{CSB in the Confining Interaction}

The existence of CSB interactions induced by mixing the electromagnetic
and strong couplings was first pointed out in Ref. 16, where it was
argued that the quark-gluon vertex picks up a charge asymmetric
contribution from photon loops\cite{15}.  Since the confining potential
must, at some level, be composed of multiple gluon exchanges\cite{16},
the existence of this coupling implies that the confining potential
will not be charge symmetric.  Unfortunately, since the confining
potential is a Lorentz scalar, the CSB contribution to the energy of
light quark hadrons is indistinguishable from the effect of quark mass
differences.

In this section, we examine the possibility of distinguishing CSB in
the confining potential from quark mass effects by looking at the
relative contribution of each to CSB in the valence quark
distribution.  Since there is, at present, no means to calculate the
confining potential, nor its correction due to charge symmetry
violation, we proceed to model the effect by altering the string
tension parameter used in the LAMP model potential. Furthermore, since
the relative normalization of the two effects is unknown, we proceed by
arbitrarily normalizing the shift in the string tension such that it
produces the same first order shift in the quark eigenvalue as the
corresponding quark mass. The wavefunctions that result from this
change are then used in Eq. 1 to produce the the charge symmetry
breaking ratios shown in Fig. 3. The solid and short-dashed curves
indicate the CSB ratios for majority and minority quarks produced by a
4 MeV quark mass difference, while the long-dashed and dot-dashed
curves are the same ratios produced by the change in the string
tension. Significantly, the CSB produced by the change in the string
tension has the opposite sign to that produced by quark mass
differences, opening the possibility that the two effects may be
distinguished from one another by precise measurement of CSB in the
valence quark distributions.

\section{Evolution}

Having separated out the CSB effects at a low momentum scale,
we must now evolve to high $Q^2$ so that a comparison with data is 
possible. Two issues arise: First, does the perturbative  evolution
erase the CSB effects we have calculated, and secondly, how large
are the CSB effects in the evolution itself? 

In Figs. 4 and 5, we show the results obtained by evolving both the CSB
pieces of the valence quark distributions and the symmetric
distributions from a low quark model scale, taken to be 0.5 GeV$^2$, to
10 GeV$^2$\cite{8}. For both majority and minority quarks the evolution
has only a small effect on the CSB ratios, shifting the curves to low
$x$ and slightly increasing the magnitude of the ratio.

Also shown in Figs. 4 and 5 is the contribution to the CSB ratio
provided by charge symmetry breaking effects generated when a quark
splits into a quark and a photon\cite{17}. To leading order in
$\alpha$, the structure of the QED and QCD contributions to the
evolution differ only by constant factors, and the effect of including
the photon diagram is to slightly speed up the rate at which the
valence distributions evolve. Since the coupling of the photon is
charge asymmetric, this means that the $u$ distribution evolves
slightly faster than the $d$ distribution. As shown in the figures,
this effect is larger than the effect of Coulomb repulsion in the quark
wavefunctions, but is nonetheless still quite small.

\section{ Comparison and Experimental Prospects}
  	
In Ref. 6, the effect of charge symmetry violation on the valence quark
distributions was calculated via the introduction of an intermediate
state diquark. The resulting two-body kinematics produces an additional
enhancement of the CSB effect produced by small changes in the diquark
mass, resulting in a 5-10\% CSB ratio for minority quarks in the range
$0.5 < x < 0.7$, roughly twice the size of our result. For majority
quarks, the two calculations  are comparable. In reference 5, using a
model independent approach, Sather obtains slightly smaller ratios for
the majority quarks than in this work, and similar results for the minority 
quarks. Generally, each of these calculations predict larger CSB effects in the
minority quark distributions than in the majority distributions.

Experimentally, the situation is less clear.  As pointed out in Ref.
18, measurements of CSB generally yield not the ratios for the
majority/minority quarks separately, but rather the ratio for the
entire valence distribution, given by
\begin{equation}
 R_{val}= {2\left [d^p_V(x)-u^p_V(x)-u^n_V(x)+d^n_V(x)\right ]\over \left [u^p_V(x)+d^p_V(x)
+u^n_V(x)+d^n_V(x)\right ]}.
\end{equation}
Since the minority quark distribution is suppressed at large $x_{Bj}$
by SU(6) symmetry breaking effects\cite{19}, the ratio is less sensitive
to the (fractionally) larger charge symmetry
breaking in the minority distribution, which is the major difference
between the models. While it is possible to isolate the minority
distribution by comparing $\pi^+-p$ and $\pi^--D$ Drell-Yan  cross
sections\cite{18}, the systematic errors associated with beam
normalizations and the deuteron EMC effect will make it difficult to
unambiguously separate the small effects predicted in this work.

Regardless of the result, the information provided by these experiments
will provide new insights into the soft dynamics of quarks in the
nucleon, and possibly a means to distinguish experimentally between CSB
generated by quark mass differences and the mixing of the strong and
electro-magnetic interactions.

\centerline{\bf Acknowledgments}
The work of C.B. and T.G. was performed under the auspices of the U.S.
Department of Energy. One of us (C.B.) gratefully acknowledges the
support of the Iowa Space Grant Cooperative and the hospitality of the
Physics Department of the Univ. of Northern Iowa during the final
stages of this work.

\pagebreak
\begin{centering}
\centerline{\underline{Figure Captions}}
\end{centering}
\begin{itemize}
\item{} Figure 1 - Charge Symmetry Ratio for Majority (valence) quarks in 
the nucleon at the quark model scale.
\item{} Figure 2 - Charge Symmetry Ratio for minority (valence) quarks in the 
nucleon at the quark model scale.
\item{} Figure 3 - Comparison of Charge Symmetry breaking produced by quark 
mass differences with that induced by mixing of the electro-magnetic and
confining interactions. 
\item{} Figure 4 - Charge Symmetry Ratio for Majority (valence) quarks in
the nucleon at a scale of 10 GeV$^2$. The sum of all contributions is
shown by the solid line, while the other lines are described in the text.
\item{} Figure 5 - Charge Symmetry Ratio for Minority (valence) quarks in
the nucleon at a scale of 10 GeV$^2$. The sum of all contributions is shown
by the solid line, while the other lines are described in the text. 
\end{itemize}

\begin{thebibliography}{99}
\bibitem{14} The issue of the consistent definition of $x_{Bj}$ is
overlooked in both Refs. 5 and 6, leading to predictions for the CSB
effect in the valence distributions which cannot, therefore, be
correct. In Ref. 6, there is a legitimate effect due to the additional
kinematic role played by the nucleon mass, but, as discussed in the
text, such effects are in disagreement with the expectations of light
cone physics. We conclude that they are artifacts of the diquark
approximation used to completely specify the model kinematics.
\end{thebibliography}
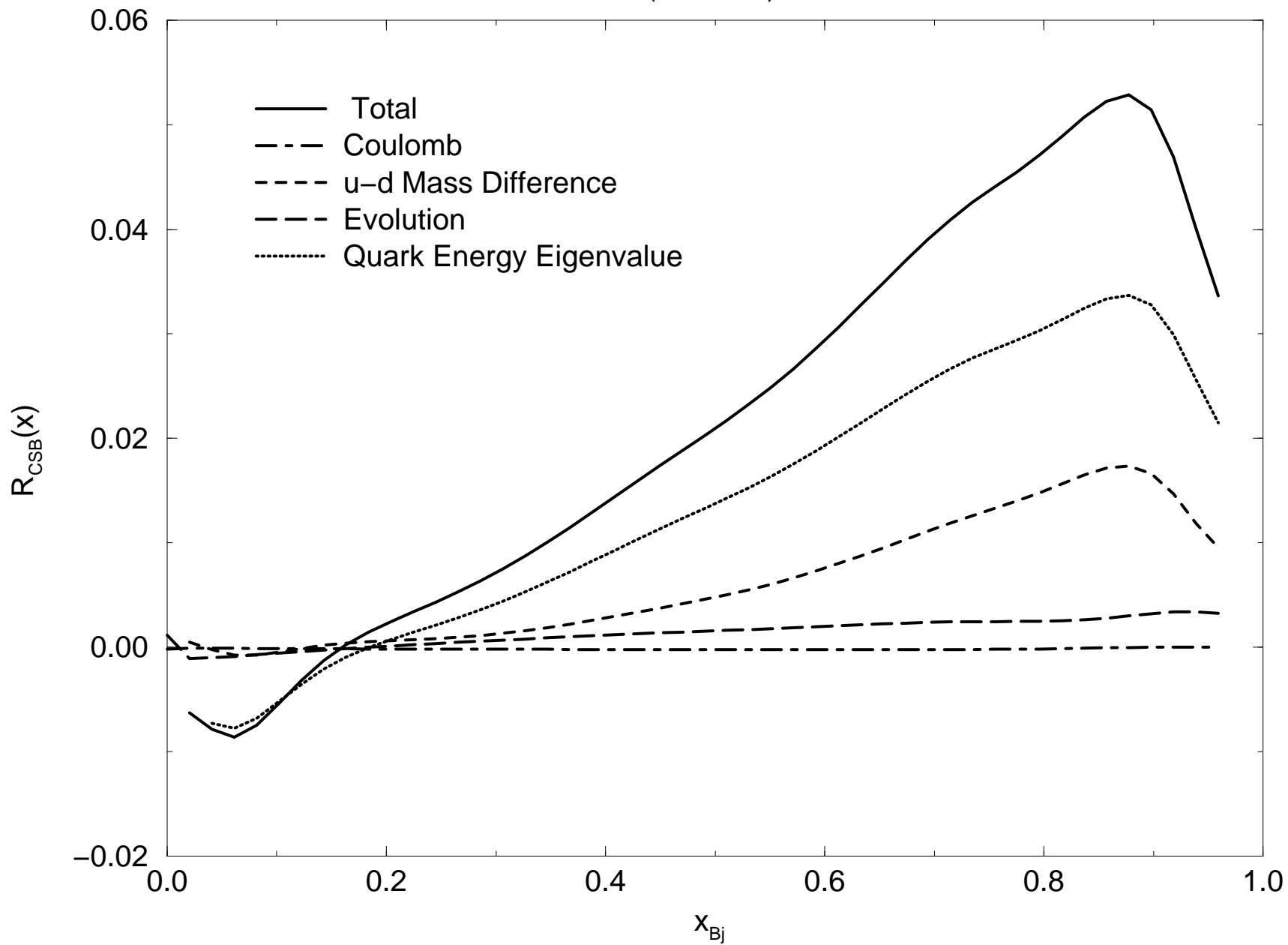
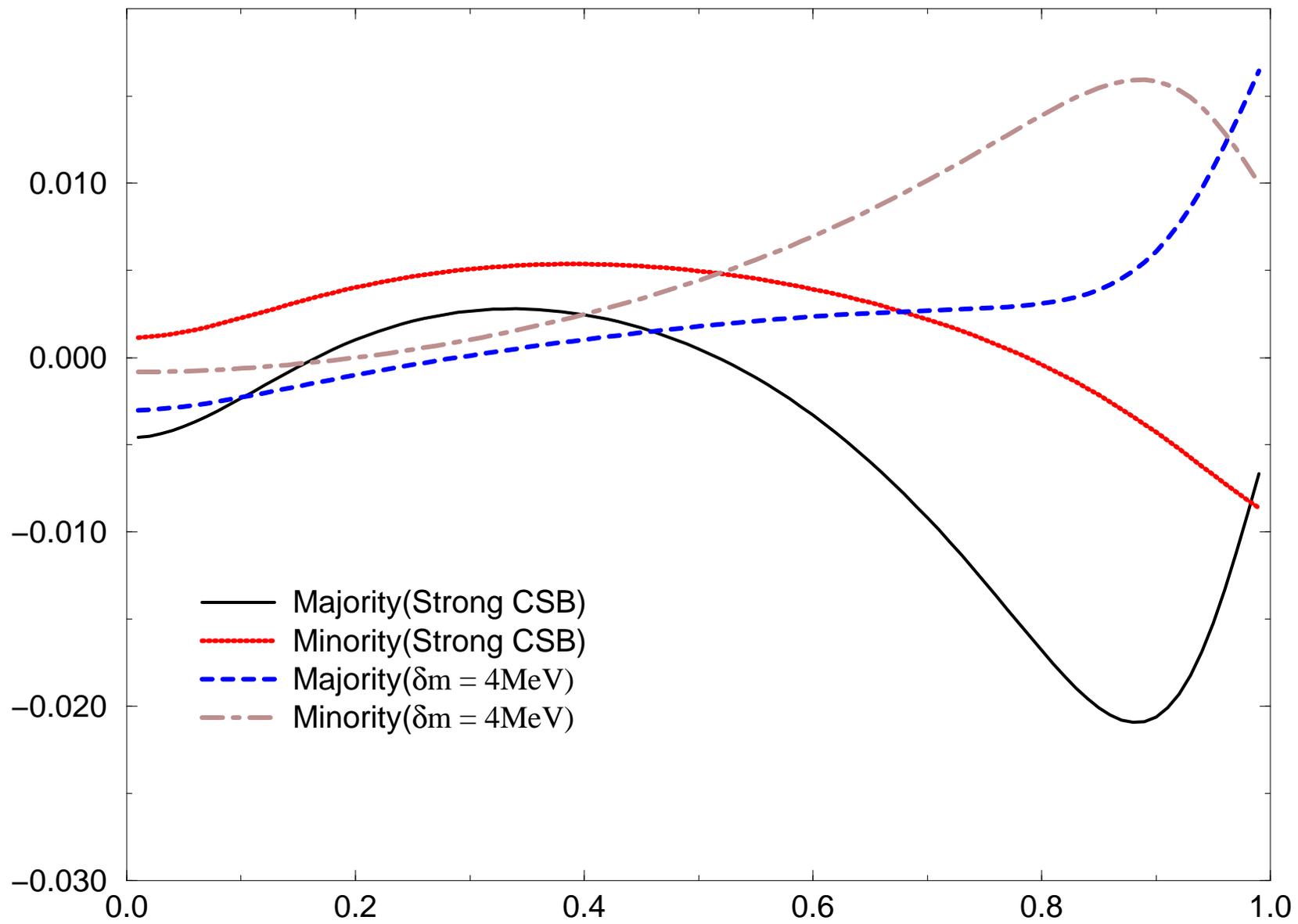
\end{document}